\documentclass[twocolumn,showpacs,preprintnumbers,amsmath,amssymb]{revtex4}
\usepackage{graphicx}% Include figure files
\usepackage{dcolumn}% Align table columns on decimal point
\usepackage{bm}% bold math

\begin{document}

\preprint{}

\title{Are photon momenta in left-handed materials reversed?}% Force line breaks with \\
\author{Hailu Luo}
%\altaffiliation[ ]{}%Lines break automatically or can be forced with \\
\author{Shuangchun Wen}\email{scwen@hnu.edu.cn}
\author{Weixing Shu}
\author{Zhixiang Tang}
\author{Yanhong Zou}
\author{Dianyuan Fan}
%%\email{Second.Author@institution.edu}
\affiliation{Key Laboratory for Micro/Nano Opto-Electronic Devices
of Ministry of Education, School of Computer and Communication,
Hunan University, Changsha 410082, China}
\date{\today}% It is always \today, today,

\begin{abstract}
We develop a semiclassical theory to describe the photon momenta in
left-handed materials (LHMs). A single two-level atom is introduced
as an ``explorer'' to probe the momenta of photons. We demonstrate
that the linear momentum of the photons reverses its direction in
LHMs. However the orbital angular momentum is remains unreversed,
although the wave-fronts reversed their screwing fashion. We
theoretically predict that the spin angular momentum is also
unreversed. The investigation of photon momenta will provide
insights into the fundamental properties of LHMs.
\end{abstract}


%Use showkeys class option if keyword
\maketitle

Left-handed materials (LHMs) would reverse many known optical
properties~\cite{Veselago1968}, such as negative
refraction~\cite{Shelby2001}, reversed Doppler
effect~\cite{Seddon2003}, negative Goos-H\"{a}nchen
shift~\cite{Berman2002}, and reversed Cherenkov
effect~\cite{Lu2003}. Vesalago theoretically predicted that the
electromagnetic momentum should reverse its direction in
LHMs~\cite{Veselago1968}. However, there exist different arguments
on the direction of linear momentum in
LHMs~\cite{Riyopoulos2006,Scalora2007,Kemp2007,Yannopapas2008}. The
issue of how to describe electromagnetic momentum is still a matter
of debate, primarily because what is considered electromagnetic and
what mechanical is to some extent arbitrary~\cite{Jackson1999}. Now
an question naturally arises: Whether is the linear momentum of
photons in LHMs reversed? In principle, the momenta can be divided
into linear momentum, orbital, and spin angular momenta. However,
the spin and orbital angular momenta have received much less
attention. Hence, we want to enquire: Whether are the spin and
orbital angular momenta also reversed?

In order to explore the momenta in LHMs, we develop a semiclassical
theory to describe the photon momentum in LHMs. Here we do not want
to involve in the famous Abraham-Minkowski controversy. The accepted
convention associates the term Abraham momentum with the quantity
${\bf D}\times{\bf B}$ and the term Minkowski momentum with the
quantity ${\bf E}\times{\bf H}/c^2$~\cite{Jackson1999}. In our
theory model, we introduce a two-level atom as an ``explorer'' to
probe the photon momenta. The interaction of  atom with a single
mode of laser beam in which the atom is treated as a quantum
two-level system and the laser field is treated as classically. It
is well known that Laguerre-Gaussian (LG) beam can carry two kinds
angular momentum: spin angular momentum of magnitude $\hbar$ per
photon due to its polarization state and orbital angular momentum of
$l\hbar$ per photon due to an azimuthal phase term $\exp[il
\varphi]$~\cite{Allen1992}. Thus, we introduce the LG beam to
describe the orbital angular momentum.

We consider a monochromatic electromagnetic field ${\bf E}({\bf
r},t) = \text{Re} [{\bf E}({\bf r})\exp(-i\omega t)]$ and ${\bf
H}({\bf r},t) = \text{Re} [{\bf H}({\bf r})\exp(-i\omega t)]$ of
angular frequency $\omega$ propagating in a homogeneous medium whose
permittivity $\varepsilon$ and permeability $\mu$ are negative
simultaneously. The field can be described by the Maxwell's
equations and the constitutive relations
\begin{eqnarray}
\nabla\times {\bf E} &=& - \frac{\partial {\bf B}}{\partial t},
~~~{\bf B} = \mu_0 \mu {\bf H},\nonumber\\
\nabla\times {\bf H} &=& \frac{\partial {\bf D}}{\partial
t},~~~~~{\bf D} =\varepsilon_0 \varepsilon {\bf E}. \label{maxwell}
\end{eqnarray}
From the Maxwell's equations, we can easily find that the wave
propagation is permitted in the medium with $\varepsilon,\mu<0$. In
this case, ${\bf E}$, ${\bf H}$, and ${\bf k}$ form a left-handed
triplet.

Under the paraxial approximation $|\partial E/\partial z|\ll k |E|$,
we obtain the wave equation
\begin{equation}
\left(\frac{\partial^2}{\partial
\rho^2}+\frac{1}{\rho}\frac{\partial}{\partial
\rho}+\frac{\partial^2}{\partial
\varphi^2}+2ik\frac{\partial}{\partial z}\right)E({\bf
\rho},\varphi,z)=0,\label{PE}
\end{equation}
where ${\bf r}=(\rho,\varphi,z)$ is cylinder polar coordinates and
$k=n\omega/c$. A particularly important solutions of the paraxial
wave equation are given by LG set of modes. In general, the LG field
in LHMs can be written as~\cite{Luo2008a}
\begin{eqnarray}
E_{pl}=&&\frac{C_{pl}}{w(z)}
\left[\frac{\sqrt{2}\rho}{w(z)}\right]^{|l|}
L_p^{|l|}\left[\frac{2\rho^2}{w^2(z)}\right]
\exp\bigg[\frac{-\rho^2}{w^2(z)}\bigg]\nonumber\\
&&\times \exp[i n k_0 z]\exp\bigg[\frac{-i n k_0 \rho^2
z}{R(z)}\bigg]\exp[ il\varphi]\nonumber\\ &&\times \exp[-i
(2p+|l|+1)\arctan (z/z_{R})],\label{Field}
\end{eqnarray}
\begin{eqnarray}
w(z)=w_0\sqrt{1+(z/z_{R})^2},~~R(z)=z+\frac{z_{R}^2}{z}.
\end{eqnarray}
Here $C_{pl}$ is the normalization constant, $L_p^{|l|}[2
\rho^2/w^2(z)]$ is a generalized Laguerre polynomial, $k_0=\omega/c$
is the wave number in vacuum, $z_{R}= n k_0 w_0^2 /2$ is the
Rayleigh length, $w(z)$ is the beam size, and $R(z)$ the radius of
curvature of the wave front. The last term denotes the Gouy phase
which is given by $\Phi=-(2p+|l|+1)\arctan(z/z_{R})$.

\begin{figure}
\includegraphics[width=7cm]{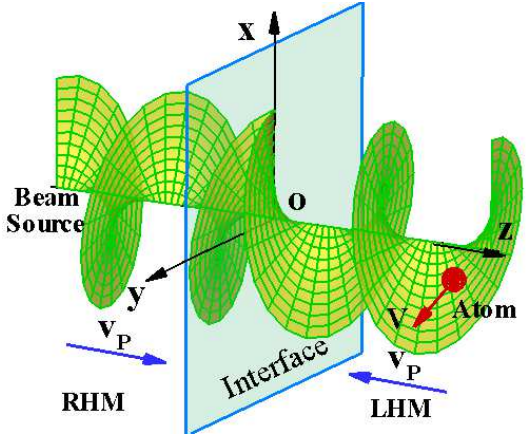}
% Here is how to import EPS art
\caption{\label{Fig1} The helical wave front for Laguerre-Gaussian
beam result from an azimuthal phase structure of $\exp[il\varphi]$
($l=+1$). The phase velocity ${\bf v}_p$ reverses its direction and
the wave-fronts reverse their screwing rotation in the LHM. When a
moving atom with velocity ${\bf V}$ across the helical wave fronts,
it experiences the dissipative force.}
\end{figure}

Laguerre-Gaussian beams with helical phase fronts, characterized by
an $\exp[il\varphi]$ azimuthal phase dependence, the orbital angular
momentum in the propagation direction has the discrete value of
$l\hbar$ per photon~\cite{Allen1992}. In the LHM, the constant
wavefront satisfies
\begin{equation}
\Theta({\bf R})=n k_0 z+\frac{-i n k_0 \rho^2
}{R(z)}+l\varphi+\Phi.\label{PP}
\end{equation}
The schematic view of the wave front is a three-dimensional screw
surface of ($\rho\cos\varphi$, $\rho \sin\varphi$, $z$). The
wavefront structure reverse their screw types with a pitch of
$\lambda_0/|n|$ (see Fig.~\ref{Fig1}). Now we want to inquire:
Whether the orbital angular momentum correspondingly reverses its
direction?

In order to probe the momenta of photons, we introduce a moving atom
as an ``explorer''. We first attempt to derive the time dependent
average force acting on the atom as a mobile center of mass
exhibiting gross motion, and the internal dynamics is described by a
two-level system interacting with laser light. The total Hamiltonian
for the atom plus field can be written as the sum of three terms
\begin{equation}
H = H_\text{field} + H_\text{atom} + H_\text{int},
\end{equation}
where $H_\text{field}$ and $H_\text{atom}$ are the zero-order
Hamiltonians for the laser and the atom, respectively. The
interaction Hamiltonian $H_\text{int}$ describes the coupling of the
atom to the electromagnetical field and is given in the electric
diploe approximation by
\begin{equation}
H_\text{int}=-{\bf d}\cdot{\bf E}({\bf R}),
\end{equation}
where ${\bf d}$ is the atomic dipole moment operator and ${\bf
E}({\bf R})$ is the electric field evaluated at position ${\bf R}$
of the atom.

\begin{figure}
\includegraphics[width=7cm]{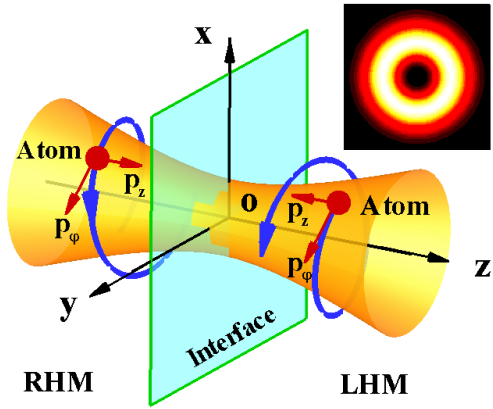}
% Here is how to import EPS art
\caption{\label{Fig2} The trajectory of a two-level atom in LG field
($l=1$) embedded in the homogeneous RHM and LHM. The liner momentum
${\bf p}_z$ reverses its direction. However the orbital angular
momentum ${\bf J}_z={\bf \rho}\times{\bf p}_\varphi$ remains its
direction unreversed. Inset: Transverse intensity distribution of
the LG field.}
\end{figure}

The average radiation force acting on the atom is defined as the
average rate of change of the atomic momentum~\cite{Allen1996}, we
can write
\begin{equation}
\langle{\bf F}({\bf R})\rangle=-\langle\nabla H_{\text{int}}\rangle.
\end{equation}
This is a time-dependent as well as spatially dependent force and it
turns out that it is divisible into two types of force, namely, the
dissipative force and a dipole force. The dissipative force
represents the force due to the absorption and reemission of the
light by the atom~\cite{Babiker1994}. To explore the linea momentum
and orbital angular momentum of photons, we want to investigate the
dissipative force, which is given by
\begin{equation}
\langle{\bf F}_{\text{diss}}({\bf R},{\bf V})\rangle=\frac{2 \hbar
\Gamma \Omega_{lp}^2({\bf R})\nabla\Theta_{lp}({\bf
R})}{\Delta_{lp}({\bf R})^2+2\Omega_{lp}({\bf R})^2+\Gamma^2},
\end{equation}
where  $\Gamma$ is the half-width of the upper quantum atomic state,
and $\nabla\Theta ({\bf R})$ is the spatial gradient of the phase of
LG beam:
\begin{eqnarray}
\nabla\Theta({\bf R})&=&\bigg[-n k_0+ \frac{n k_0
\rho^2}{2(z^2+z_R^2)}\left(\frac{2z^2}{z^2+z_R^2}-1\right)\nonumber\\
&&-\frac{(2p+|l|+1)z_R}{z^2+z_R^2}\bigg]{\bf e}_z+\frac{n k_0
\rho^2}{R(z)}{\bf e}_\rho+ \frac{l}{\rho}{\bf e}_\varphi.\label{F1}
\end{eqnarray}
The plane-wave phase emerges directly from by setting $l=0$, $p=0$,
and $z_R\rightarrow \infty$. The function $\Omega_{pl}({\bf R})$ is
identified as the position-dependent Rabi frequency:
\begin{equation}
\Omega_{pl}({\bf R})=\frac{\Omega_{00} C_{pl}}{1+z^2/z_R^2}
\left[\frac{\sqrt{2}\rho}{w(z)}\right]^{|l|} L_p^{|l|}\left[\frac{2
\rho^2}{w^2(z)}\right] \exp\bigg[\frac{-\rho^2}{w^2(z)}\bigg],
\end{equation}
which is well defined for a given plane-wave Rabi frequency
$\Omega_{00}$ and LG field. The dynamic detuning is defined as
\begin{equation}
\Delta_{pl}({\bf R})=\Delta_0-\nabla\Theta_{lp}({\bf R})\cdot{\bf
V},
\end{equation}
where $\Delta_0=\omega-\omega_0$ is the static detuning, with
$\hbar\omega_0$  the level energy separation of the two-level atom
and $\omega$ the frequency of the light. Lagurre-Gaussian field
exerts momenta to the atom is given by ${\bf F}_{\text{diss}}=d {\bf
P}/dt$. To explore whether the photon momentum is reversed, we need
to judge the direction of dissipative force. The force of the atom
experiencing in a LG field is given by
\begin{eqnarray}
\langle{\bf F}_{\text{diss}}({\bf R})\rangle&=&\frac{\hbar \Gamma
\mathcal{J}}{1+\mathcal{J}+\Delta^2/\Gamma^2}\bigg\{-\frac{n k_0
\rho}{R} {\bf e}_\rho-\frac{l}{\rho}{\bf e}_\varphi\nonumber\\&&+
\bigg[-n k_0+ \frac{n k_0
\rho^2}{2(z^2+z_R^2)}\left(\frac{2z^2}{z^2+z_R^2}-1\right)\nonumber\\
&&-\frac{(2p+|l|+1)z_R}{z^2+z_R^2}\bigg]{\bf e}_z
\bigg\},\label{Dop}
\end{eqnarray}
where $\mathcal{J}=2G^2({\bf R})/\Gamma^2$ is the position-dependent
saturation parameter. The motion of the atom governed by ${\bf
F}_{\text{diss}}=d^2 {\bf R}(t)/dt^2$, and the trajectory is
depicted in Fig.~\ref{Fig2}. The dominant component of the
dissipative force is the axial term:
\begin{equation}
\langle{\bf F}_{\text{axial}}\rangle=\frac{n k_0 \hbar \Gamma
\mathcal{J}}{1+\mathcal{J}+\Delta^2/\Gamma^2} {\bf e}_z.
\end{equation}
Note that the wave vector reverse its direction, thus the axial
dissipative force should reversed its direction. Thus the linear
momentum of photon ${\bf p}_z=\hbar {\bf k}$ is reversed, which
demonstrates Veselago's early prediction~\cite{Veselago1968}.

The azimuthal is the only force that is response for a torque on the
atom about the propagating axis of LG beam,
\begin{equation}
\langle{\bf F}_{\text{azimuth}}\rangle=\frac{\hbar \Gamma
\mathcal{J}}{\Delta^2+\Gamma^2+2\Omega_{lp}^2({\bf R})}\frac{l
}{\rho}{\bf e}_\varphi.
\end{equation}
The torque is given by $\langle{\bf T}\rangle=\langle{\bf
\rho}\times{\bf F}_\text{azimuth}\rangle$. Clearly, the atom will
rotate anticlockwise ($l>0$) or clockwise ($l<0$) about the beam
axis with a angular velocity. This can be easily understood if we
consider the torques under the condition of saturation limit
$\mathcal{J}\rightarrow\infty$, the torque deduces to the simple
form
\begin{equation}
\langle {\bf T}\rangle\approx l\hbar \Gamma {\bf e}_z.
\end{equation}
The is proportional to the orbital angular momentum of photon
$l\hbar$, which is independent of the refractive index. The atom
still remains its rotation fashion in the LHM (see Fig.~\ref{Fig2}).
Obviously, the orbital angular momentum of photon in LHMs is
unreversed. This interesting feature is consistent with the
appetence of the unreversed rotational Doppler
effect~\cite{Luo2008b}.

Note that the light remains linearly polarized and the intrinsic
spin plays no role in our theory model. When a birefringent particle
such as a calcite fragment is introduced, and circularly polarized
light would be converted to linear. In principle, the spin angular
momentum should cause a particle to spin about its own axis. The
sense of rotation is governed by the spin angular momentum of
photons~\cite{He1995,Neil2002}. In our opinion, spin and orbital
angular momenta do not depend upon the refractive index and so is
said to be intrinsic. Hence we predict that the spin angular
momentum should remain its direction unchanged. Further research is
needed to demonstrate whether the spin angular momentum is reversed.
It is possible that the study of photon momenta in LHMs may make a
useful contribution to long established Abraham-Minkowski dilemma.

In summary, we have developed a semiclassical theory to describe the
photon momenta in LHMs. A single two-level atom has been introduced
to probe the momenta of photons. We have demonstrated that the
linear momentum in LHMs should be reversed. However the orbital
angular momentum of photons remains unreversed, although the
wave-fronts reversed their screwing fashion. We predict that not all
optical effects would reverse their properties in LHMs. Some
intrinsic optical effect, such as spin and orbital angular momenta
of photon, should unreverse its direction. The semiclassical theory
can be applied to explore other intriguing phenomena in LHMs, such
as photon recoil, Doppler effect, and photon drag effect. To explore
whether these effect are reversed would allow us to better
understand the interaction of light with LHMs.

We wish to acknowledge the support of projects of the National
Natural Science Foundation of China (Grants Nos. 10674045, 10804029,
50802027, and 60538010).

\end{document}